\documentclass[12pt]{article}
\usepackage{epsfig}
\newcommand{\be}{\begin{equation}}
\newcommand{\ee}{\end{equation}}
\newcommand{\bqa}{\begin{eqnarray}}
\newcommand{\eqa}{\end{eqnarray}}
\newcommand{\pslash}{\slash\hspace{-0.55em}}
\newcommand{\as}{\alpha_{\mathrm{s}}}

\begin{document}
\begin{center}
{\LARGE Problems of QCD factorization in exclusive decays of $B$ meson to charmonium}\\[0.8cm]
{\large Zhongzhi Song$^{(a)}$ and Kuang-Ta Chao$^{(b,a)}$}\\[0.5cm]
{\footnotesize (a)~Department of Physics, Peking University,
 Beijing 100871, People's Republic of China}\\
{\footnotesize (b)~China Center of Advanced Science and Technology
(World Laboratory), Beijing 100080, People's Republic of China}
\end{center}
\vspace{0.5cm}

\begin{abstract}
We study the exclusive decays of $B$ meson into P-wave charmonium
states $\chi_{cJ}(J=0,1)$ in the QCD factorization approach with
light-cone distribution functions describing the mesons in the
processes. For $B \rightarrow \chi_{c1} K$ decay, we find that
there are logarithmic divergences arising from nonfactorizable
spectator interactions even at twist-2 order and the decay rate is
too small to accommodate the experimental data. For $B\rightarrow
\chi_{c0} K$ decay, we find that aside from the logarithmic
divergences arising from spectator interactions at leading-twist
order, more importantly, the factorization will break down due to
the infrared divergence arising from nonfactorizable vertex
corrections, which is independent of the specific form of the
light-cone distribution functions. Our results may indicate that
QCD factorization in the present form may not be safely applied to
$B$-meson exclusive decays to charmonium states.
 \vspace{1cm}\\
PACS numbers: 13.25.Hw; 14.40.Gx
\end{abstract}

\vspace{1cm} Exclusive nonleptonic $B$-meson decays provide a
important opportunity to determine the parameters of the
Cabibbo-Kobayashi-Maskawa (CKM) matrix, to explore CP violation
and to observe new physics effects. Recently, $B$ physics has
received extensive experimental attention  such as from
high-energy experiments at the Tevatron and at the $e^+ e^-~B$
factories. On the other hand, although the underlying weak decay
of the $b$ quark is simple, quantitative understanding of
nonleptonic $B$-meson decays is difficult due to the complicated
strong-interaction effects.

Beneke et al. have considered two-body nonleptonic $B$-meson
decays extensively including light-light as well as heavy-light
final states within the QCD factorization
approach\cite{BBNS1,BBNS2,BBNS3}. The general idea is that in the
heavy quark limit $m_b \gg \Lambda_{\mathrm{QCD}}$, the transition
matrix elements of operators in the hadronic decay $B\to M_1M_2$
with $M_1$ being the recoiled meson and $M_2$ being the emitted
meson can be calculated in an expansion in $1/m_b$ and $\as$. The
leading term in $1/m_b$ assumes a simple form\cite{BBNS2}:
 \bqa \label{qcdf}
\langle M_1M_2|O_i|B\rangle\!=\!F^{BM_1}(m_2^2)\int^1_0
\!du\,T^I(u)\phi_{M_2}(u)\!+\!\int^1_0 \!d\xi \,du\,dv
\,T^{II}(\xi,u,v)\phi_B(\xi)\phi_{M_1}(v)\phi_{M_2}(u),
 \eqa
where $M_2$ is a light meson or a quarkonium and $F^{BM_1}$ is the
$B \to M_1$ transition form factor; $\phi_M$ is meson light-cone
distribution amplitude and $T^{I,II}$ are perturbatively
calculable hard scattering kernels. If we neglect strong
interaction corrections, eq.(\ref{qcdf}) reproduces the result of
naive factorization. However, hard gluon exchange between $M_2$
and $BM_1$ system implies a nontrivial convolution of hard
scattering kernels $T^{I,II}$ with the light-cone distribution
amplitude $\phi_{M_2}$. This method works well for light-light
final states\cite{BBNS1,BBNS3,PP,PV} as well as heavy-light final
states\cite{BBNS2,DL}.

Exclusive $B$-meson decays to charmonium are important since those
decays e.g. $B \to J/\psi K$ are regarded as the golden channels
for the study of CP violation in $B$ decays. It is argued that
because the size of the charmonium is small$(\sim 1/{\as
{m_\psi}})$ and its overlap with the $(B, K)$ system may be
negligible\cite{BBNS2}, the same QCD factorization method as for
$B\to\pi\pi$ can be used for $B \to J/\psi K(K^*)$ decay. Indeed,
explicit calculations\cite{chay, cheng} show that the
nonfactorizable vertex contribution is infrared safe and the
spectator contribution is perturbatively calculable, where the
light-cone distribution functions are used for $B$, $K$ as well as
$J/\psi$ mesons. This small size argument for the applicability of
QCD factorization for charmonia is intuitive and interesting, but
it needs verifying for charmonium states e.g. the P-wave
$\chi_{cJ}$ states other than the $J/\psi$. In addition, recently
BaBar and Belle collaborations have measured the exclusive decays
of $B \rightarrow \chi_{cJ} K (J=0,1)$\cite{babar, belle1}. So, it
is also interesting to compare the predictions based on the QCD
factorization approach with the experimental data. In this letter,
we report the problems of the QCD factorization approach
encountered in these two decay channels. As in \cite{chay,cheng},
in the following we will use light-cone distribution functions to
describe $B$, $K$, as well as charmonium mesons.

We first consider $\overline{B}\rightarrow \chi_{c1} K$ decay. The
effective Hamiltonian is written as\cite{BBL}
 \bqa
H_{\mathrm{eff}} = \frac{G_F}{\sqrt{2}} \Bigl( V_{cb} V_{cs}^*
(C_1 {\cal O}_1 +C_2 {\cal O}_2 ) -V_{tb} V_{ts}^* \sum_{i=3}^{10}
C_i {\cal O}_i \Bigr),
 \eqa
where $G_F$ is the Fermi constant, $C_i$ are the Wilson
coefficients and $V_{q_1q_2}$ are the CKM matrix elements. Here
the relevant operators ${\cal O}_i$ are given by
  \bqa
&&\,\,{\cal O}_1=(\overline{s}_{\alpha} b_{\beta})_{V-A} \cdot
(\overline{c}_{\beta} c_{\alpha})_{V-A},\qquad\qquad\qquad~ {\cal
O}_2=(\overline{s}_{\alpha} b_{\alpha})_{V-A} \cdot
(\overline{c}_{\beta} c_{\beta})_{V-A},
 \nonumber\\
&& {\cal O}_{3(5)}=(\overline{s}_{\alpha} b_{\alpha})_{V-A} \cdot
\sum_q (\overline{q}_{\beta} q_{\beta})_{V-A(V+A)},\qquad~ {\cal
O}_{4(6)}=(\overline{s}_{\alpha} b_{\beta})_{V-A} \cdot \sum_q
(\overline{q}_{\beta} q_{\alpha})_{V-A(V+A)},
\\
&& {\cal O}_{7(9)}={3\over 2}(\overline{s}_{\alpha}
b_{\alpha})_{V-A} \cdot \sum_q e_q (\overline{q}_{\beta}
q_{\beta})_{V+A(V-A)},~ {\cal O}_{8(10)}={3\over
2}(\overline{s}_{\alpha} b_{\beta})_{V-A} \cdot \sum_q e_q
(\overline{q}_{\beta} q_{\alpha})_{V+A(V-A)},\nonumber
 \eqa
where $\alpha,~\beta$ are color indices and the sum over $q$ runs
over $u, d, s, c$ and $b$. Here $(\bar q_1 q_2)_{V\pm A}=\bar
q_1\gamma_\mu (1\pm\gamma_5) q_2 $.

To calculate the decay amplitude, we introduce the $\chi_{c1}$
decay constant $f_{\chi_{c1}}$ as\cite{ball1}
 \be\label{vector2}
 \langle \chi_{c1} | \overline{c}
(0) \gamma_{\mu}\gamma_5 c(0) |0\rangle = -if_{\chi_{c1}}
m_{\chi_{c1}}\epsilon^*_{\mu},\\
\ee where $m_{\chi_{c1}}$ is the mass of $\chi_{c1}$ and
$\epsilon$ is the $\chi_{c1}$ polarization vector. The decay
constant $f_{\chi_{c1}}$ is a nonperturbative quantity and may be
estimated from the potential models, the QCD sum rules, or lattice
QCD calculations.

The leading-twist light-cone distribution amplitude of $\chi_{c1}$
can be expressed as
 \bqa  \label{lcone}
 \langle \chi_{c1}
(p,\epsilon)| \overline{c}_{\alpha} (y) c_{\beta} (z) |0\rangle
=\frac{i}{4} \int_0^1 du \cdot e^{i(u p\cdot y + (1-u) p \cdot z)}
f_{\chi_{c1}} m_{\chi_{c1}} \bigl( \pslash{\epsilon}^* \gamma_5
\bigr)_{\beta \alpha} \phi_{\chi_{c1}}(u),
 \eqa
where $u$ and $1-u$ are respectively the momentum fractions of the
$c$ and $\bar c$ quarks inside the $\chi_{c1}$ meson, and the wave
function $\phi_{\chi_{c1}}(u)$ for $\chi_{c1}$ meson is symmetric
under $u\leftrightarrow 1-u$.

In naive factorization, we neglect the strong interaction
corrections and the power corrections in
$\Lambda_{\mathrm{QCD}}/m_b$. Then the decay amplitude at leading
order is written as \bqa \label{tree2} iM_0=2i  f_{\chi_{c1}}
m_{\chi_{c1}} \epsilon^*\cdot p_B F_1 (m_{\chi_{c1}}^2)
\frac{G_F}{\sqrt{2}}\Bigl[ V_{cb} V_{cs}^* (C_2 +\frac{C_1}{N_c}
)-V_{tb} V_{ts}^* (C_3 + \frac{C_4}{N_c} -C_5 -\frac{C_6}{N_c})
\Bigr],
 \eqa
where $p_B$ is the momentum of $B$ meson, $F_1$ is the
$B\rightarrow K$ transition form factor and $N_c$ is the number of
colors. We do not include the effects of the electroweak penguin
operators since they are numerically small\cite{BBL}.

The form factors for $\overline{B} \rightarrow K$ are given as
  \bqa \label{vmu2}
\langle K(p_K) | \overline{s} \gamma_{\mu} b| B(p_B)\rangle=\Bigl[
(p_B +p_K)_{\mu} -\frac{m_B^2-m_K^2}{p^2} p_{\mu} \Bigr] F_1 (p^2)
+ \frac{m_B^2-m_K^2}{p^2} p_{\mu} F_0 (p^2),
  \eqa
where $p= p_B -p_K$ is the momentum of $\chi_{c1}$ with $p^2 =
m_{\chi_{c1}}^2$, $m_B$ and $m_K$ are respectively the masses of
$B$ and $K$ mesons. We will neglect the kaon mass for simplicity.
We can use the ratio between these two form factors as\cite{chay}
\bqa \frac{F_0(p^2)}{F_1 (p^2)} = 1-\frac{p^2}{m_B^2}.
 \eqa
 So we need only one of the two form factors to describe the decay amplitude.

\begin{figure}[t]
\begin{center}
\vspace{-2.7cm}
\includegraphics[width=14cm,height=14cm]{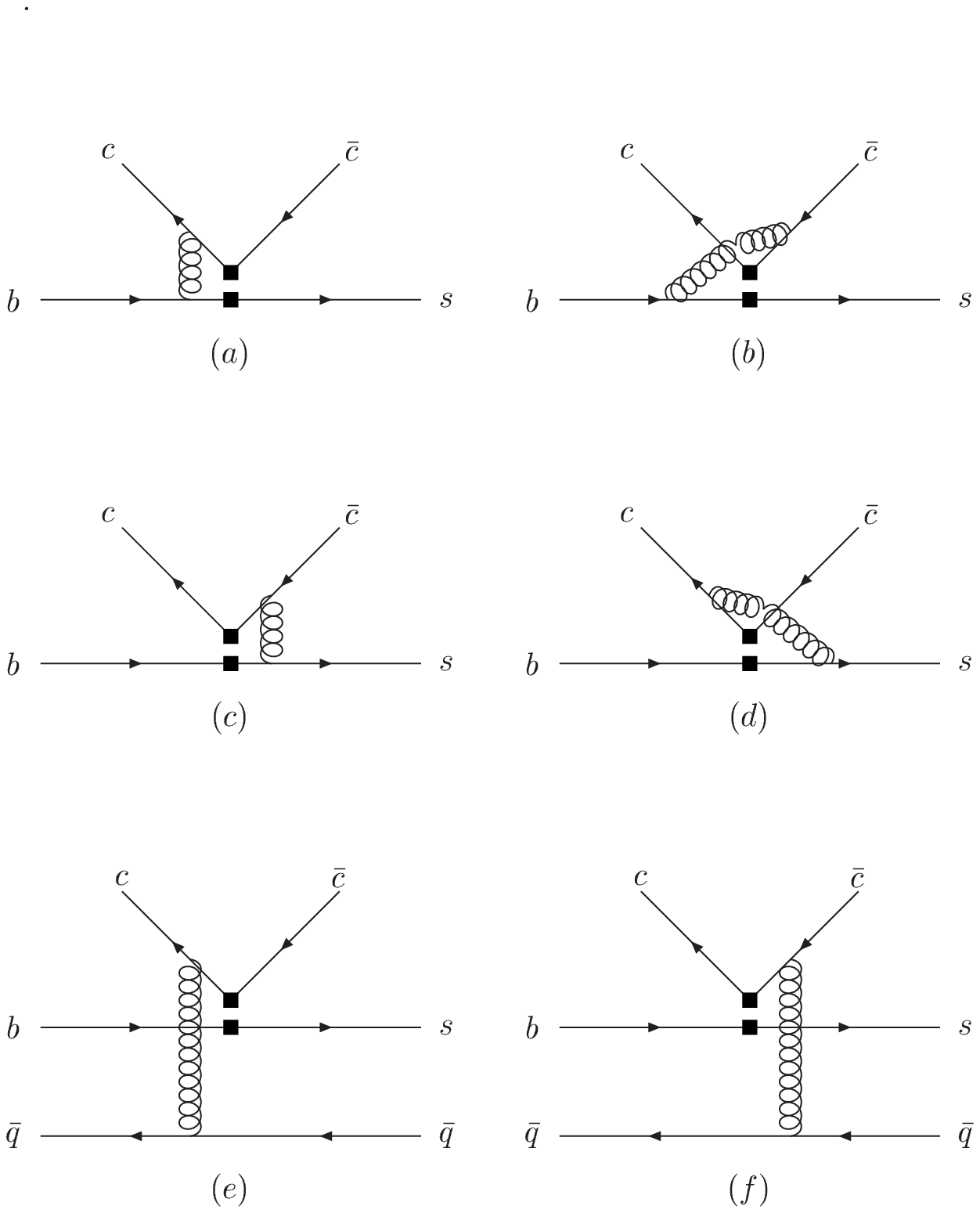}
\vspace{-4.2cm}
\end{center}
\caption{ Feynman diagrams for vertex and spectator corrections to
$\overline{B} \rightarrow \chi_{c1} K$.} \label{fvs}
\end{figure}

Taking nonfactorizable corrections in Fig.\ref{fvs} into account,
the full amplitude for $ \overline{B} \rightarrow \chi_{c1} K$ in
QCD factorization is written compactly as
 \bqa\label{total2}
  iM = 2i f_{\chi_{c1}} m_{\chi_{c1}} \epsilon^*
\cdot p_B F_1 (m_{\chi_{c1}}^2) \frac{G_F}{\sqrt{2}} \Bigl[V_{cb}
V_{cs}^* a_2 -V_{tb} V_{ts}^* (a_3 - a_5) \Bigr],
 \eqa
where the coefficients $a_i$ ($i=2,3,5$) in the naive dimension
regularization(NDR) scheme are given by  \bqa\label{ai}
a_2=C_2+\frac{C_1}{N_c}+\frac{\alpha_s}{4\pi}\frac{C_F}{N_c} C_1
\Bigl(-18 +12\ln \frac{m_b}{\mu} + f_I + f_{II}
\Bigr), \nonumber \\
a_3=C_3+\frac{C_4}{N_c}+\frac{\alpha_s}{4\pi}\frac{C_F}{N_c} C_4
\Bigl(-18 +12\ln \frac{m_b}{\mu}+f_I + f_{II} \Bigr),
 \\
a_5=C_5+\frac{C_6}{N_c}-\frac{\alpha_s}{4\pi}\frac{C_F}{N_c} C_6
\Bigl(-6 +12\ln \frac{m_b}{\mu} +f_I + f_{II}\Bigr),\nonumber
 \eqa
 where $C_F=(N_c^2-1)/(2 N_c)$ and $m_b$ is the $b$ quark mass.

The function $f_I$ is calculated from the four vertex correction
diagrams (a, b, c, d) in Fig.\ref{fvs} and reads
 \bqa
  f_I\!=\!\int_0^1 du \phi_{\chi_{c1}} (u)
  \Bigl[\frac{3(1-2u)}{1-u} \ln u -3i \pi +3 \ln (1-z)
  + \frac{z^2 (1-u)^2 \Bigl(\ln (1-z)
-i\pi\Bigr)}{( 1-z u)^2}
\nonumber \\
+\frac{2z (1-u)}{1-z u}+\Bigl( \frac{1-u}{(1-z u)^2}
-\frac{u}{\bigl(1-z (1-u)\bigr)^2} \Bigr) z^2 u \ln (z u)\Bigr],
 \eqa
where $z={m^2_{\chi_{c1}}}/{m^2_B}$, and we have already
symmetrized the result with respect to $u\leftrightarrow 1-u$.

The function $f_{II}$ is calculated from the two spectator
correction diagrams(e, f) in Fig.\ref{fvs} and it is given by
 \bqa\label{fII}
f_{II}=\frac{4\pi^2}{N_c} \frac{f_K f_B}{F_1 (m_{\chi_{c1}}^2)
m_B^2} \frac{1}{1-z} \int_0^1 d\xi \frac{\phi_B (\xi)}{\xi}
\int_0^1 dy \frac{\phi_K (y)}{y} \int_0^1 du
{\phi_{\chi_{c1}}(u)}[\frac{1}{u}+\frac{2z}{y(1-z)}],
 \eqa
 where $\xi$ is the momentum fraction of the
spectator quark in the $B$ meson and $y$ is the momentum fraction
of the $s$ quark inside the $K$ meson, $\phi_B$, $\phi_K$ are the
light-cone wave functions for the $B$ and $K$ meson respectively.
$f_{II}$ corresponds to the $T^{II}$ term in Eq.(\ref{qcdf}).

The spectator contribution depends on the wave function $\phi_B$
through the integral
 \bqa
\int_0^1 d\xi \frac{\phi_B(\xi)}{\xi} \equiv
\frac{m_B}{\lambda_B}.
 \eqa
 Since $\phi_B (\xi)$ is appreciable only for $\xi$ of order
$\Lambda_{\mathrm{QCD}}/m_B$, $\lambda_B$ is of order
$\Lambda_{\mathrm{QCD}}$. We will choose $\lambda_B\approx 300$
MeV in the numerical calculations\cite {BBNS3}.

There is an integral related to $\phi_K (y)$ in Eq.(\ref {fII})
which will give logarithmic divergence. Therefore QCD
factorization breaks down even at leading order. This is different
from $B \rightarrow J/\psi K$ decay which does not have
logarithmic divergence at leading twist\cite{chay, cheng}. The
reason is that the logarithmic divergences arising from
contributions of vector and tensor currents are cancelled out in
the $B \rightarrow J/\psi K$ decay, whereas there is no such
cancellation for the $B \rightarrow \chi_{c1} K$ decay. Following
Ref.\cite {BBNS3}, we treat the divergent integral as an unknown
parameter and write
  \be\label{xh}
\int_0^1 dy \frac{\phi_K (y)}{y^2}=\int_0^1 dy
\frac{6(1-y)}{y}=6(X_H-1),
 \ee
where the asymptotic form $\phi_K (y)=6 y (1-y)$ is used for the
kaon twist-2 light-cone distribution amplitude. And we will also
choose $X_H=\ln(m_B/\Lambda_{QCD})$\cite {BBNS3} as a rough
estimate in our calculation.

For numerical analysis, we choose $F_1 (m_{\chi_{c1}}^2) =
0.743$\cite {ball2} and use the following input parameters:
 \bqa
m_b=4.8~\mbox{GeV},  \ \ m_B=5.28~\mbox{GeV}, \ m_{\chi_{c1}}=3.5
~\mbox{GeV},  \nonumber\\
f_{\chi_{c1}}=230~ \mbox{MeV}, \ f_B = 180 ~ \mbox{MeV}, \  f_K =
160 ~\mbox{MeV}.
 \eqa
Here for the $\chi_{c1}$ decay constant $f_{\chi_{c1}}$ we give
its expression in the heavy quarkonium potential model as \bqa
 f_{\chi_{c1}}=\sqrt{\frac{18}{\pi m_c^2
m_{\chi_{c1}}}}R'(0),
 \eqa
where $R'(0)$ is the derivative of the $\chi_{c1}$ radial wave
function at the origin, and we may roughly estimate its value by
using some potential models, e.g. given in Ref.\cite{quig}.

\begin{table}[t]
\begin{center}
\begin{tabular}{ c | c c c c c c }
   \hline
    &$C_1$ & $C_2$ & $C_3$ & $C_4$
 & $C_5$ & $C_6$ \\
 \hline
    NDR & 1.082 & -0.185 & 0.014 & -0.035 & 0.009 & -0.041 \\
 \hline
 \end{tabular}
\caption{ {Next-to-leading-order Wilson coefficients in NDR scheme
(See Ref.\cite{BBL}) with $\mu=4.4$ GeV and
$\Lambda^{(5)}_{\overline{\rm MS}}=225$ MeV.}}
 \label {wilson}
\end{center}
\vspace{-0.5cm}
\end{table}

\begin{table}[tb]
\begin {center}
\begin{tabular}{ c|ccc}
 \hline
 $\phi_{\chi_{c1}}(u)$ &$a_2$&$a_3$&$a_5$ \\   \hline
$6u(1-u)$&0.1255-0.0815i&0.0040+0.0026i&-0.0027-0.0031i \\
$\delta (u-1/2)$&0.1154-0.0814i&0.0043+0.0026i&-0.0031-0.0031i \\
\hline
\end{tabular}
\caption{The coefficients $a_i$ at $\mu=4.4$ GeV with different
choices of $\phi_{\chi_{c1}}(u)$.}
 \label{table2}
\end {center}
\vspace{-0.5cm}
\end{table}

The specific form of the $\chi_{c1}$ light-cone distribution
amplitude is not known, but as in the $J/\psi$ case (see
\cite{chay,cheng}) we may use the light meson asymptotic form
$\phi_{\chi_{c1}}(u) = 6u (1-u)$ as an approximate description. In
the numerical analysis, we also consider the form
$\phi_{\chi_{c1}}(u) =\delta (u-1/2)$, which comes from the naive
expectation corresponding to the nonrelativistic limit of the
distribution amplitude. Although there are uncertainties
associated with the form of the distribution function, we will see
shortly that the calculated decay rates are not sensitive to the
choice of the distribution amplitude.

Using the next-to-leading-order (NLO) Wilson coefficients in
Table.~\ref {wilson}, we get the results of coefficients $a_i$
which are listed in Table.~\ref {table2}. With the help of these
coefficients $a_i$, we calculate the decay branching ratios with
two different choices of the $\chi_{c1}$ distribution amplitude
and get
 \bqa
{\mathrm{Br}} (\overline{B} \rightarrow \chi_{c1} K) = 0.97 \times
10^{-4},\,\, ( \phi_{\chi_{c1}} (u) = 6u(1-u));\\
{\mathrm{Br}} (\overline{B} \rightarrow \chi_{c1} K) = 0.87 \times
10^{-4}, \,\,( \phi_{\chi_{c1}}(u) = \delta (u-1/2)). \eqa

 The measured branching ratio is\cite{babar}
  \be
{\mathrm{Br}} (B^0 \rightarrow \chi_{c1} K^0) = (5.4 \pm 1.4)
\times 10^{-4},
  \ee
which is about six times larger than the  theoretical results.

We now consider the $B \rightarrow \chi_{c0} K$ decay. Unlike the
$\chi_{c1}$ mode that we just analyzed, $\chi_{c0}$ does not have
couplings to $V$ or $A$ currents, so this decay mode is prohibited
in naive factorization. But it may occur if there is an exchange
of an additional gluon\cite{beneke, diehl}. The branching ratio
from experiment is\cite{belle1}
 \be
{\mathrm{Br}} (B^+ \rightarrow \chi_{c0} K^+) = (6.0 \pm 2.1)
\times 10^{-4}.
 \ee

Similar to the calculation performed above, we write the
$\chi_{c0}$ light-cone distribution amplitude in the following
general form
 \bqa \label{lcda0}
  \langle
\chi_{c0} (p)| \overline{c}_{\alpha} (y) c_{\beta} (z) |0\rangle =
\frac{i}{4} \int_0^1 du e^{i(u p\cdot y + (1-u) p \cdot z)}
\Bigl[f_{\chi_{c0}}^s m_{\chi_{c0}}\textsf{1}_{\beta \alpha}
\phi_{\chi_{c0}}^s(u) + f_{\chi_{c0}}^v \bigl( \pslash{p}
\bigr)_{\beta \alpha} \phi_{\chi_{c0}}^v(u)\Bigr],
 \eqa
where $f_{\chi_{c0}}^s, f_{\chi_{c0}}^v, \phi_{\chi_{c0}}^s(u)$
and $\phi_{\chi_{c0}}^v(u)$ denote the $\chi_{c0}$ decay constants
and light-cone distribution amplitudes, respectively for the
scalar and vector currents. It is easy to see that for the scalar
meson $\chi_{c0}$ the decay constants for the vector current and
the antisymmetric tensor current all vanish (i.e., $\langle
\chi_{c0}| \overline{c} \gamma_\mu c|0\rangle=\langle \chi_{c0}|
\overline{c} \sigma_{\mu\nu} c|0\rangle=0$), and only the scalar
current decay constant is nonzero. Therefore on the right hand
side of Eq.(\ref{lcda0}) only the first term makes a real
contribution and it is essentially the leading twist (twist-3)
contribution. However, here we also list the second term (twist-2)
in Eq.(\ref{lcda0}) in order to show that it would also give an
infrared divergence contribution if the vector current decay
constant were nonzero.

Because of charge conjugation invariance,$~\phi_{\chi_{c0}}^s(u)$
is symmetric while $~\phi_{\chi_{c0}}^v(u)$ is anti-symmetric
under $u \leftrightarrow 1-u$. The Feynman diagrams of order $\as$
correction for $B \rightarrow \chi_{c0} K$ are the same as those
in Fig.\ref{fvs}. In the calculation of $\chi_{c1}$, the
contribution of the four vertex diagrams in Fig.~\ref {fvs} is
infrared safe. However, we find there are infrared divergences
arising from the vertex diagrams in the $B \rightarrow \chi_{c0}
K$ decay. Keeping only the divergent terms, we get
 \bqa\label{V}
V_{Div}=\frac{\alpha_s C_F}{\pi N_c} i F_1(m_{\chi_{c0}}^2)\cdot
\Bigl\{ m_c m_{\chi_{c0}} f_{\chi_{c0}}^s \int_0^1
\phi_{\chi_{c0}}^s(u) du \Bigl[ -{1\over u}\ln{\lambda} \ln{z}
\Bigr] +m_B^2 f_{\chi_{c0}}^v \int_0^1 \phi_{\chi_{c0}}^v(u) du
\nonumber \\
\times\Bigl[ -\ln{\lambda} \ln{z}+(1-z)[\ln{\lambda}
 +{1\over 2} \ln^2(\frac{\lambda u z}{1-z})
- i \pi \ln(\frac{\lambda u z}{1-z})] \Bigr] \Bigr \},
 \eqa
where $z=m_{\chi_{c0}}^2/ m_B^2$ and $\lambda$ corresponds to the
soft gluon momentum cutoff divided by the charm quark mass. The
infrared divergence is explicitly seen as $\lambda \to 0$.

The contribution arising from spectator interactions is given by
 \bqa \label{H}
  H=\frac{\alpha_s C_F \pi}{ N_c^2} i f_K f_B \Bigl[
  \frac{ m_c m_{\chi_{c0}}}{m_B^2-m_{\chi_{c0}}^2}f_{\chi_{c0}}^s
 \int_0^1 d\xi \frac{\phi_B (\xi)}{\xi}
 \int_0^1 dy \frac{\phi_K (y)}{{y}^2}
 \int_0^1 du \frac{\phi_{\chi_{c0}}^s(u)}{u (1-u)}
\nonumber\\
+f_{\chi_{c0}}^v\int_0^1 d\xi \frac{\phi_B (\xi)}{\xi}\int_0^1 dy
\frac{\phi_K (y)}{{y}} \int_0^1 du
\frac{\phi_{\chi_{c0}}^v(u)}{1-u}\Bigr].
 \eqa
There are logarithmic divergences even in the leading-twist order.
So, soft gluon exchange dominates $B \rightarrow \chi_{c0} K$
decay.

In the above calculations, we have chosen the heavy quark limit as
$m_b\to \infty$ with $m_c/m_b$ fixed. In this limit, the
logarithmic divergences in Eq.(\ref {fII}) for $B \rightarrow
\chi_{c1} K$ decay, and the infrared divergences in Eq.(\ref V) as
well as the logarithmic divergences in Eq.(\ref H) for $B
\rightarrow \chi_{c0} K$ decay, will break down QCD factorization.
Another choice of the heavy quark limit is that $m_b\to \infty$
with $m_c$ fixed. Then all the divergences mentioned above are
power corrections and should be dropped out, so QCD factorization
still holds in this limit. Physically, the latter case is
equivalent to the limit of zero charm quark mass in which
charmonium is regarded as a light meson. Obviously, the first
choice of the heavy quark limit with $m_c/m_b$ fixed is more
relevant in phenomenological analyses, and is also usually used in
theoretical studies \cite{BBNS2, BBNS4}, where it is expected that
QCD factorization should apply to $B$ meson exclusive decays into
charmonium in the limit $m_c\rightarrow\infty$ with corrections of
order $\Lambda_{\rm QCD}/(m_c\alpha_s)\sim 1$. Our result shows
that this expectation holds for decays like $B \to J/\psi K$ but
not for e.g. $B \to \chi_{c0} K$ where the vertex infrared
divergence will break down factorization at order of $\Lambda_{\rm
QCD}/(m_c\alpha_s)$. It is also worthwhile to emphasize that the
infrared divergences in Eq.(\ref V) are more serious than the
logarithmic divergences in Eqs.(\ref {fII}),(\ref H). This is
because, when one includes the parton transverse degrees of
freedom, end-point singularities arising from spectator
interactions can be regularized within the framework of $k_T$
factorization \cite{kt} and logarithmic divergences can then be
removed. However, the infrared divergences due to nonfactorizable
vertex corrections still exist, and they are independent of the
specific form of the light-cone distribution functions.

In summary, we have studied the exclusive decays of $B$ meson into
P-wave charmonium states $\chi_{cJ}(J=0,1)$ in the QCD
factorization approach with light-cone distribution functions
describing mesons involved in the processes. For $B \rightarrow
\chi_{c1} K$ decay, the factorization breaks down due to
logarithmic divergences arising from nonfactorizable spectator
interactions at twist-2 order and the decay rate is too small to
accommodate the data. The situation for $B\rightarrow \chi_{c0} K$
decay is even more serious because of the infrared divergence
arising from nonfactorizable vertex corrections as well as
logarithmic divergences due to spectator interactions at the
leading-twist order. Considering the above problems encountered in
$B\rightarrow \chi_{cJ}K$ decays and especially the infrared
divergence in $B\rightarrow \chi_{c0}K$ decay, we intend to
emphsize that the QCD factorization method with the present form
may not be safely applied to exclusive decays of $B$ meson into
charmonium, e.g. P-wave states. However, it should be noted that
our results are obtained in QCD factorization by using the
light-cone distribution functions for charmonium (as in e.g.
refs.\cite{chay,cheng}), which may not be a very appropriate
description for heavy quarkonium. To further clarify the problem,
many theoretical studies such as use of the nonrelativistic QCD
for charmonium, the perturbative QCD factorization method, and the
nonperturbative QCD effects should be attempted. In any case, a
better understanding is definitely needed for describing $B$-meson
exclusive decays to charmonium, and for reducing the big gap
between the present theoretical results and the experimental
observations.

\section*{Acknowledgements}
We are grateful to H.-n. Li for helpful discussions on this work.
We also thank J.P. Ma, C.S. Huang, D.X. Zhang and D.S. Yang for
useful discussions and comments. We would like to thank C. Meng
for checking the calculations. This work was supported in part by
the National Natural Science Foundation of China, and the
Education Ministry of China.

\end{document}